\documentclass[twocolumn,pra,showpacs]{revtex4}

\usepackage{amsmath,amssymb}
\usepackage{graphics}
\usepackage{epsfig}

\begin{document}

\title{Enhancing the Observability of the Efimov Effect in Ultracold
Atomic Gas Mixtures} 
\author{J. P. D'Incao and B. D. Esry}
\affiliation{Department of Physics, Kansas State University,
Manhattan, Kansas 66506} 

\begin{abstract}
We discuss the prospects for observing the characteristic features
of the Efimov effect in a two-component ultracold atomic
gas near an interspecies Feshbach resonance. In the ultracold regime,
the Efimov effect is expected to be manifested in the three-body
collision rates through the appearance of series of minima or maxima
as a function of the two-body $s$-wave scattering length $a$. 
Here, we propose the observation of this Efimov physics through
measurements of the inelastic three-body rate constants 
near a Feshbach resonance.
Our analysis suggests that boson-fermion mixtures, where
the bosons are much heavier than the fermions, are the
most favorable system to observe such features. 
\end{abstract} 
\pacs{34.10.+x,32.80.Cy,05.30.Jp}
\maketitle 

The Efimov effect was originally predicted in nuclear physics in the
early 1970's \cite{Efimov}. The Efimov effect is the emergence of a 
large number of three-body bound states even when the two-body
subsystems have none, and it requires that at least two of the
interparticle interactions be resonant.  Besides this remarkable
effect on the three-body bound state spectrum, it has been shown 
\cite{ScaLenDep,MassDep,BraatenReview} that Efimov physics has
far-reaching influence on ultracold three-body collisions which, in
turn, can be a primary atomic and molecular loss mechanism in
ultracold quantum gases. To date, however, it exists only as a
theoretical prediction, since it has not been conclusively
demonstrated experimentally.    

The ability to experimentally control the interatomic interaction 
by tuning an external magnetic field near a Feshbach resonance in
ultracold atomic gases \cite{Ketterle-FR,Wieman-FR,Verhaar-FR} creates
new possibilities for observing  
Efimov physics. Near such a resonance, the two-body $s$-wave scattering
length $a$ can vary from $-\infty$ to $+\infty$, revealing Efimov
physics in the three-body system whenever $|a|$ is much larger than
the characteristic size of the interatomic interactions $r_{0}$. In
this regime, ultracold three-body observables behave universally in
the sense that their dependence on $a$ can be predicted while the
details of the interatomic interactions enter only through a small
number of short range parameters. The clearest manifestation of 
Efimov physics is through the appearance of a
series of minima or maxima in the three-body observables that are
equally spaced on a $\ln |a|$ scale. The multiplicative factor that
separates these features is  $e^{\pi/s_{0}}$, where $s_{0}$ depends on
the mass ratio between the collision partners and the number of
resonant interactions. Such series have been predicted for three-body
collision processes such as three-body recombination,
collision-induced dissociation, and vibrational relaxation
\cite{ScaLenDep,MassDep,BraatenReview}.   

While direct experimental evidence of the Efimov {\em effect} is not 
available, some recent measurements of three-body collision rates can
be interpreted as indirect evidence for Efimov {\em
physics}. Specifically, the predicted $a^4$  
scaling law for recombination of identical bosons
\cite{Ketterle-K3,Grimm-K3} has been observed as has the $a^{-3.33}$
suppression of the relaxation of weakly bound molecules in two-spin
fermionic gases \cite{Salomon-LLM,Grimm-LLM,Hulet-LLM,Jin-LLM}. 
Despite these successes and the experimental control now available,
the most convincing signature of Efimov physics, namely equally
spaced features on a $\ln |a|$ scale, has not yet been observed.  
Furthermore, such observations are unlikely to be made in the near
future for the equal mass systems most studied theoretically since the
features are expected to occur each time $a$ increases by
$e^{\pi/s_0}$$\approx$22.7. To experimentally verify that their
spacing is equal, a minimum of three such features must be observed.
Since theory can only predict their spacing and not their positions,
$a$ must experimentally cover a range of about 22.7$^4$, which is not
easily achievable. 

In this paper we argue that measuring a sequence of three features
arising from Efimov physics can be made much more tractable by simply
changing the constituents of the ultracold gas. In particular,
boson-fermion mixtures \cite{Ketterle-BF,Jin-BF}, where the bosonic
atoms are much heavier than the fermionic atoms, exhibit the most
favorable conditions for observing Efimov physics. Our proposal is
to measure the inelastic loss rates through measurements of the time
evolution of both atom and molecule numbers near a Feshbach resonance,
as was done, for instance, in
Refs.~\cite{Ketterle-K3,Grimm-K3,Jin-LLM,Jin-BF}.  
In general, the use of mixtures of heavy and light atoms leads to more
favorable conditions for observing Efimov physics
\cite{MassDep}. In this limit, the space between the minima or maxima
is smaller and, therefore, our goal of three Efimov features can be
achieved with a more experimentally manageable range of $a$. This
statement applies equally well to fermion-fermion and boson-boson
mixtures, but other conditions complicate using these systems. We also
present numerical calculations that indicate how other issues,
especially the presence of several two-body bound states, might 
limit the observability of Efimov physics in ultracold mixtures
of commonly used alkali atoms. 

In two-component ultracold atomic gases --- of, say, $X$ atoms and $Y$
atoms --- only two systems involve interspecies interactions:
$XXY$ and $XYY$.  Ideally, both species would be prepared in hyperfine
states that minimize two-body losses so that three-body collisions
become dominant. In this case, extracting the inelastic three-body
rates would then be cleaner, requiring monitoring the atomic 
and molecular densities over time.   
The rate equations governing this evolution, however, depend on
contributions from both $XXY$ and $XYY$ systems. For instance, the
rate equation for the atomic density $n_{X}$
depends on the three-body recombination rate constant for $X+X+Y$ and
$X+Y+Y$ collisions and the vibrational relaxation rate for $XY+X$
collisions. Similarly, the rate equation for $n_{Y}$ depends on the 
$X+X+Y$, $X+Y+Y$, and $XY+Y$ collision rates, while 
the rate equation for 
$n_{XY}$, where $XY$ are weakly bound molecules, depends on the $XY+X$
and $XY+Y$ collisions, as well as relaxation due molecule-molecule
collisions, $XY+XY$. 
Near a Feshbach resonance, however, the
importance of each collision process for the atomic and molecular
densities evolution is dictated by its energy and scattering length
dependence \cite{ScaLenDep,MassDep} and, evidently, on the product of
densities of each specie involved in the collision process.
In describing the rate equations, we have assumed that the
intraspecies interactions are not resonant, that two-atom processes
are not important, that collision-induced dissociation is not
energetically allowed, and that all inelastic processes result in
loss.   

To observe Efimov physics through three-body recombination ($K_{3}$),
no molecules should be trapped so that recombination correlates
directly to atomic loss. In other words, the atomic densities $n_{X}$
and $n_{Y}$ are affected solely by $X+X+Y$ and $X+Y+Y$ collisions.
%In other words, $n_{XY}\approx 0$ so that
%$V_{\rm rel}$ drops out of Eqs.~(\ref{nX}) and (\ref{nY}). 
Tracking the atomic densities should thus reveal Efimov physics. If
a substantial number of weakly bound molecules are also trapped, then
Efimov physics should be observed through vibrational relaxation
($V_{\rm rel}$), monitoring both the atomic and molecular 
densities. In this case, relaxation due to molecule-molecule
collisions must also be considered \cite{Petrov}, but no Efimov
features have yet been predicted for such collisions.

We have previously obtained \cite{ScaLenDep,MassDep} the energy and
scattering length dependence for the ultracold three-body rate
constants and found that processes exhibiting minima due to Efimov
physics are modulated by the factor 
\begin{equation}
M_{s_{0}}(a)=\sin^{2}\left[s_{0}\ln({a}/{r_{0}})+\Phi\right]
\label{Minima}
\end{equation}
and that processes for which Efimov physics emerges as peaks share the
factor 
\begin{equation}
P_{s_{0}}(a)=\frac{\sinh(2\eta)}{\sin^2\left[s_0\ln({|a|}/{r_{0}})+\Phi\right]+\sinh^2(\eta)}. 
\label{Peaks}
\end{equation}
In the above equations, $\Phi$ is an unknown phase and $\eta$
parametrizes the probability of making a transition to a deeply bound
state --- both arise from short range physics associated with details
of the interatomic interactions and are expected to differ for each
system.  It is from these expressions that the spacing factor
$e^{\pi/s_0}$ is derived.   

\begin{figure}[htbp]
\includegraphics[width=1.9in,angle=270]{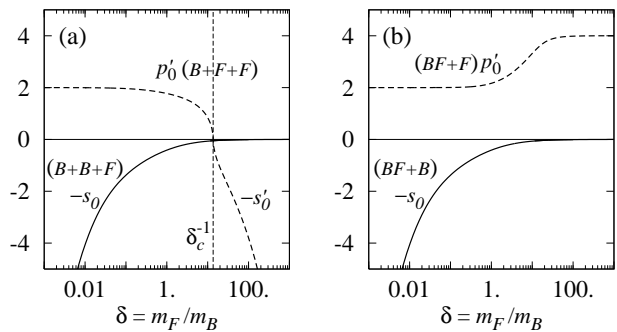}
\caption{Coefficients $s_{0}$ and $p_{0}$ as a function of the 
  mass ratio $\delta$ for (a) three-body recombination and
  (b) vibrational relaxation for boson-fermion mixtures.
\label{FigCoeffs}}  
\end{figure}

Since one of the main hindrances to observing Efimov physics in a
``traditional'' equal mass boson system is the need to experimentally
vary $a$ by a factor of roughly $(e^{\pi/s_0})^4$$\approx$$22.7^4$
($s_0$$\approx$1.0064) in order to see three successive minima or
maxima, one way to simplify the experiment is to reduce the spacing
between the features.  It is clear that this can only be accomplished
by changing $s_0$.  As it turns out \cite{Efimov,ScaLenDep,MassDep},
$s_0$ depends on the masses of the constituent
atoms. Figure~{\ref{FigCoeffs}} shows this dependence for both $BBF$ 
and $BFF$ systems as a function of the mass ratio $\delta$=$m_F/m_B$.   
The influence of the identical particle
symmetry is seen in the difference between the curves labeled $s_0$
($BBF$) and $s_0'$ ($BFF$).  Since we can realistically expect atomic
masses that differ by at most a factor of 20--30, $s_0$ can be
increased to roughly two --- a dramatic improvement since the spacing
is determined by an exponential.

Figure~\ref{FigCoeffs} also shows the mass dependence of the parameter 
$p_0$ \cite{Efimov,ScaLenDep,MassDep}.  This parameter is also
influenced by Efimov physics and is the key parameter determining the
suppression of vibrational relaxation in $FX+F$ collisions
\cite{MassDep}. Even though $p_0$ is influenced by Efimov physics,
rate constants that depend on it do not show minima or maxima as do
rate constants that depend on $s_0$. We note, though, that even the
$BFF$ systems can show Efimov features. In
Fig.~\ref{FigCoeffs}(a), the $BFF$ curve is labeled with $p_0'$ at
small $\delta$, but transitions to $s_0'$ at $\delta$ larger than
$\delta_c^{-1}$ (defined this way for consistency with
Ref.~\cite{MassDep}). Recombination in $BFF$ systems will thus only
show minima or maxima for $\delta$ larger than 13.607, spaced by
$e^{\pi/s_0'}$.  
\begin{table}[htbp]
\caption{Scattering length dependence of the three-body
  rate constants in boson-fermion ($B$-$F$)
  mixtures. Boldface indicates the processes
  expected to exhibit the Efimov effect, characterized by a
  series of minima ($M_{s_{0}}$) or maxima ($P_{s_{0}}$) given,
  respectively, by   Eqs.~(\ref{Minima}) and (\ref{Peaks}). 
}
\label{TabRatesMixtures}
\begin{ruledtabular}
\begin{tabular}{ccccc}
  \multicolumn{5}{c}{$K_{3}$} \\
   \multicolumn{2}{c}{{$B+B+F$}} &
 & \multicolumn{2}{c}{{$B+F+F$} ($\delta<\delta_{c}^{-1},~
  \delta>\delta_{c}^{-1}$)} \\
 \cline{1-2} \cline{3-5}
   $a>0$ 
 & $a<0$ &
 & $a>0$
 & $a<0$  \\ \hline 
   \boldmath{$M_{s_{0}}a^4$} 
 & \boldmath{$P_{s_{0}}|a|^4$} &
 & $Ea^6$,\boldmath{$EM_{s'_{0}}a^6$} 
 & $E|a|^{6-2p'_{0}}$, \boldmath{$EP_{s'_{0}}|a|^{6}$} \\
  \multicolumn{5}{c}{$V_{\rm rel}$} \\ 
   \multicolumn{2}{c}{{$BF+B$}} &
 & \multicolumn{2}{c}{{$BF+F$}} \\ 
\cline{1-2} \cline{3-5}
   $a>0$
 & $a<0$ &
 & $a>0$
 & $a<0$ 
\\ \hline
 \boldmath{$P_{s_{0}}a$} 
 & const &
 & $a^{1-2p'_{0}}$      
 & const 
\end{tabular}
\end{ruledtabular}
\end{table}

Table~\ref{TabRatesMixtures} summarizes the scattering length and
energy dependence of all of the three-body rate constants relevant to
mixtures of bosons and fermions. The cases in bold are expected to
exhibit Efimov physics either through minima [$M_{s_0}$ from
Eq.~(\ref{Minima})] or maxima [$P_{s_0}$ from Eq.~(\ref{Peaks})]. A 
word of caution is in order at this point.  The scaling behavior
listed in the table holds only so long as the system is in the
threshold regime, i.e, when the collision energy is the smallest
energy \cite{Limits}. Since most experiments with ultracold atoms are
done at essentially fixed collision energy (temperature), the
requirement of being in the threshold regime defines the maximum value
of the scattering length for which threshold behavior can be expected.
Explicitly, $a_{\rm max}=\hbar/(2\mu_{XY}E)^{1/2}$, where $\mu_{XY}$
is the two-body reduced mass.  For $a>a_{\rm max}$ the threshold
scaling laws break down and finite energy effects quickly wash out the
features related to Efimov physics. As a consequence, even at
ultracold temperatures only a finite number of minima or maxima can
be expected to be observed. 

%Equations~(\ref{nX})--(\ref{nXY}) indicate that, 
In general, the observability of Efimov physics depends not only on
the rate constants, but also on the various atomic and molecular
densities. This is certainly the case for boson-boson and
fermion-fermion mixtures. For recombination in boson-fermion mixtures,
however, the exact values of the atomic densities are not crucial
since $B+F+F$ collisions are suppressed by a factor of the energy at 
ultracold temperatures. $B+B+F$ collisions will thus dominate, making 
boson-fermion mixtures the top candidate for observing Efimov 
physics. Furthermore, choosing $m_{B}\gg m_{F}$ makes $s_0$ large and 
puts $BFF$ recombination in the regime without minima or maxima. 
Of course, as mentioned before, molecules should be neither present
nor trapped so that atomic loss is due mainly to 
recombination in $B+B+F$ collisions.  Taking all of these points into 
consideration, the rate equations are approximately  
\begin{equation}
{\dot{n}_{B}}\approx\dot{n}_F\approx -K^{B+B+F}_{3}n_{B}^2n_{F},
\end{equation}
\noindent
and minima and peaks (Table~\ref{TabRatesMixtures}) should be
observed for $a>0$ and $a<0$, respectively. 

If substantial numbers of molecules are present in the gas, the
density dependence in the rate equations favors vibrational relaxation
as the best process to measure for effects of Efimov physics.
In particular, since relaxation for $BF+F$ collisions is suppressed as
$a^{1-2p_0'}$ [2$\leq p_0'\leq$4 from Fig.~\ref{FigCoeffs}(b)],
measuring peaks in the $BF+B$ relaxation constant will reveal Efimov
physics. In addition, molecule-molecule collisions are also suppressed
since $BF$ molecules are composite fermions, eliminating $s$-wave
collisions by symmetry. The time evolution for both atomic and
molecular species is given by 
\begin{eqnarray}
{\dot{n}_{B}}&\approx&{\dot{n}_{BF}}\approx-V^{BF+B}_{\rm
  rel}n_{BF}n_{B}\nonumber \\ 
{\dot{n}_{F}}&\approx&-V^{BF+F}_{\rm rel}n_{BF}n_{F}. 
\end{eqnarray}
\noindent
In this case, Efimov physics is manifested only for $a>0$
(Table~\ref{TabRatesMixtures}) due to $BF+B$ collisions.

\begin{figure}[htbp]
\includegraphics[width=2.35in,angle=270]{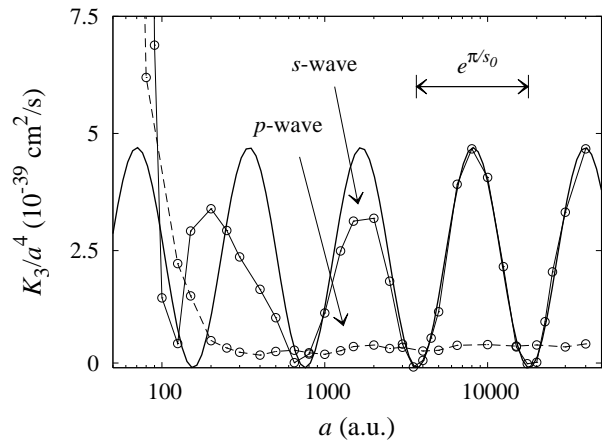}
\caption{Recombination for model $^{133}$Cs+$^{133}$Cs+$^{6}$Li
  collisions into weakly bound $s$-wave $^{133}$Cs$^{6}$Li molecules 
  (solid line with circles) and in to deeply bound $p$-wave
  $^{133}$Cs$^{6}$Li molecules (dashed line with 
  circles). The solid line represents the analytical expression
  [Eq.(\ref{Minima})].}\label{FigK3BBF}   
\end{figure}

Figure~\ref{FigK3BBF} shows numerical results for  $a>0$ recombination
using a model system with masses corresponding to
$^{133}$Cs+$^{133}$Cs+$^{6}$Li collisions, disregarding the same
species interactions. We have calculated the rates for scattering
lengths up to $5\times10^4$~a.u. at a collision energy of $1$~nK by
using a two-body model potential adjusted to support two $s$-wave and
one $p$-wave bound states, and assuming $r_{0}=15$~a.u.. In  
Figure~\ref{FigK3BBF} we show recombination to weakly bound $s$-wave
$^{133}$Cs$^{6}$Li molecules and to deeply
bound $p$-wave $^{133}$Cs$^{6}$Li molecules 
--- recombination to deeply bound $s$-wave was about two orders of
magnitude smaller than recombination to deeply bound $p$-wave
molecules. In this range of $a$, we found four minima 
with spacing given by $e^{\pi/s_{0}}=4.88$
($s_{0}=1.982766$).   
For large $a$, the numerical results confirm the
analytical formula from Table~\ref{TabRatesMixtures}.
Not unexpectedly, the results for small $a$ differ from the analytical  
formula as details of the interatomic interactions become
important. As a result, the first minimum in Fig.~\ref{FigK3BBF} does 
not follow the $e^{\pi/s_{0}}$ spacing. The expected spacing is
recovered, though, between the second and higher minima.

Figure~\ref{FigK3BBF} also shows that even in the presence of
recombination to deeply bound states, the contrast ratio of the
oscillations should be measurable.
Further, since recombination to highest vibrationally
excited channels usually dominates, we expect this result to hold even
for more realistic potentials with many two-body bound
states. Nevertheless, recombination to deeply bound states could set
limits on the observation of Efimov physics by washing out the minima
or reducing the contrast. Preliminary numerical calculations suggest,
however, that a more realistic two-body potential suppress
recombination to deeply bound states that shown in Fig.~\ref{FigK3BBF}.

\begin{table}[htbp]
\caption{Values for the spacing $e^{\pi/s_{0}}$ between the Efimov
  features expected in $B+B+F$ and $BF+B$ collisions in boson-fermion
  mixtures. For each mixture, $a_{\rm min}$ and $E_{\rm max}$ indicate
  the estimated requirements for observing at least four Efimov
  features.\label{TabII}}    
\begin{ruledtabular}
\begin{tabular}{rccc}
 & \multicolumn{3}{c}{{$K_{3}^{B+B+F}$ and $V_{\rm rel}^{BF+B}$}}  \\
  \cline{2-4} 
\multicolumn{1}{c}{$B-F$} & $e^{\pi/s_{0}}$ & $|a_{\rm min}|$(a.u.)
& $E_{\rm max}$(nK)  \\\hline 
$^{133}$Cs$-$$^{6}$Li & $4.877$ & $1.6\times10^4$ & $60.0$  \\
 $^{87}$Rb$-$$^{6}$Li & $6.856$ & $5.6\times10^4$ & $5.00$    \\
 $^{23}$Na$-$$^{6}$Li & $36.28$ & $3.3\times10^8$ & $\ll0.1$ \\
  $^{7}$Li$-$$^{6}$Li & $>100$ & $\gg10^8$       & $\ll0.1$ \\
$^{133}$Cs$-$$^{40}$K & $47.02$ & $9.2\times10^7$ & $\ll0.1$ \\
 $^{87}$Rb$-$$^{40}$K & $>100$ & $\gg10^8$       & $\ll0.1$ \\
 $^{23}$Na$-$$^{40}$K & $>100$ & $\gg10^8$       & $\ll0.1$ \\
  $^{7}$Li$-$$^{40}$K & $>100$ & $\gg10^8$       & $\ll0.1$ 
\end{tabular}
\end{ruledtabular}
\end{table}

Table~\ref{TabII} summarizes the prospects for observing the second, 
third, and fourth Efimov features in boson-fermion mixtures of 
commonly used alkali atoms (assuming the first feature does not follow
predictions, as shown in Fig.~\ref{FigK3BBF}).  
The table shows the expected spacing of Efimov features given the
masses of the atoms, the minimum magnitude $a_{\rm min}$ to which $a$ 
must be tuned, and the maximum energy for which the system remains
in the threshold regime for $a$ up to $a_{\rm min}$. The minimum $a$
is determined from $a_{\rm min}=r_{0}e^{N\pi/s_{0}}/f(\delta)$, with
$N$=4 and $f(\delta)=[\sqrt{\delta(\delta+2)}/(\delta+1)]^{1/2}$
\cite{MassDep}; and the maximum energy, from $E_{\rm
  max}=\hbar^2/2\mu_{BF}a_{\rm min}^2$. In Table~\ref{TabII} we have
assumed $r_{0}=15$~a.u. for all mixtures and thus the table gives only
a rough estimate for $a_{\rm min}$ and $E_{\rm max}$. 
Table~\ref{TabII} includes boson-fermion mixtures used in recent
experiments: 
$^{23}$Na-$^{6}$Li~\cite{Ketterle-BF} and
$^{87}$Rb-$^{40}$K~\cite{Jin-BF}, as well as
$^{7}$Li-$^{6}$Li~\cite{Kokkelmans}. 
Not surprisingly, the systems with the largest mass ratios,
$^{133}$Cs-$^{6}$Li and 
$^{87}$Rb-$^{6}$Li, provide the most favorable prospects for observing
Efimov physics since both mixtures offer manageable temperatures and a
reasonable range of $a$.      

Before closing, we briefly consider observing Efimov physics in other
systems. In boson-boson mixtures, Efimov physics is observable in
$K_{3}$ and $V_{\rm rel}$ \cite{MassDep}. Unlike boson-fermion
mixtures, however, none of the collision rates are suppressed in the
ultracold regime.  Thus, the competition might wash out
the features related to Efimov physics. The situation is similar in 
fermion-fermion mixtures. In this case, though, Efimov physics is
visible only in $K_{3}$ for mixtures of heavy and light fermions
\cite{MassDep} and is suppressed by a factor of $E$ at ultracold
temperatures. In both cases, the control of atomic and molecular
densities would be crucial for observing Efimov physics.

In this paper we have discussed the prospects for experimentally
observing Efimov physics in two-component ultracold quantum gases by
measuring the inelastic loss rates near a Feshbach resonance. We have
shown that boson-fermion mixtures with heavy bosons and light
fermions are probably most likely to exhibit the
signatures of Efimov physics. Both the range of $a$ and the maximum
allowed temperature are experimentally accessible, allowing a
sufficient number of Efimov features to be observed to establish their
equal spacing in $\ln(a)$. These results suggest that a first
experimental demonstration of the Efimov effect is, in fact, currently
possible.    

This work was supported by the National Science Foundation.

\end{document}